\documentclass[fleqn,twoside]{article}
\usepackage{espcrc2}
\usepackage{amsmath}
\usepackage{graphicx}
\usepackage{here}
\usepackage[figuresright]{rotating}

\title{Performance of the AMS-02 Transition Radiation Detector}
\author{
         Ph.v.~Doetinchem,
         S.~Fopp, 
         W.~Karpinski, 
        Th.~Kirn, 
     	 K.~L\"ubelsmeyer, 
     	 J.~Orboeck, 
     	 S.~Schael,
     	 A.~Schultz von Dratzig, 
     	 G.~Schwering, 
        Th.~Siedenburg,
     	 R.~Siedling, 
     	 W.~Wallraff \\
       {\em I. Physikalisches Institut RWTH-Aachen} \\
     	 U.~Becker, 
     	 J.~Burger, 
     	 R.~Henning, 
     	 A.~Kounine, 
     	 V.~Koutsenko, 
         J. Wyatt\\
       {\em Massachusetts Institute of Technology, Cambridge} \\
  }

\begin{document}

\begin{abstract}

For cosmic particle spectroscopy on the International Space Station
the AMS experiment will be equipped with a Transition Radiation Detector (TRD)
to improve particle identification.
The TRD has 20 layers of fleece radiator with Xe/CO$_2$ proportional-mode straw-tube chambers.
They are supported in a conically shaped octagon structure made of CFC-Al-honeycomb.
For low power consumption VA analog multiplexers are used as front-end readout.
A 20 layer prototype built from final design components has achieved proton rejections 
from 100 to 2000 at 90\% electron efficiency for proton beam energies up to 250 GeV 
with cluster counting, likelihood and neural net selection algorithms.

\vspace{1pc}
\end{abstract}

\maketitle

\section{Introduction}
\label{section1}
The AMS-02 experiment will measure the cosmic ray particle spectra on the 
International Space Station ISS for a period of at least three years.   

\begin{figure}[H]
\hspace{-0.5cm}
\includegraphics[width=8.0cm,clip]{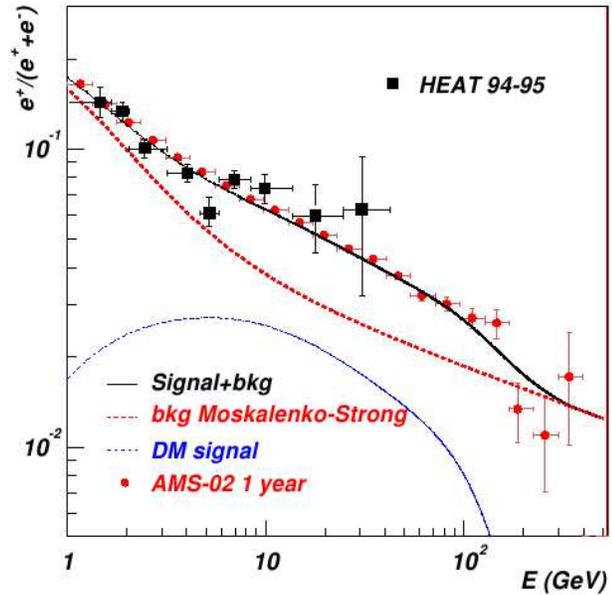}
\caption{\it The expected positron fraction accuracy from AMS-02 after 1 year on the ISS compared to available data 
from the HEAT collaboration \cite{HEAT}. 
The error bars reflect the particle identification power estimated from AMS subdetector beamtest data \cite{AMS02_Construction}.}
\label{AMS02_pos_frac}
\end{figure}

With AMS-02 data our understanding of dark matter could be substantially
improved. There is overwhelming evidence that some non-luminous non-baryonic substance
of yet unknown nature accounts for more than 80\% of the matter density of the universe
\cite{WMAP,WMAP_SDSS_SN1a}.
The neutralino $\chi$ as a weakly interacting massive particle
is the most promising candidate. It would be indirectly observable as additional
signal from neutralino annihilation processes in positron spectra
as shown in fig. \ref{AMS02_pos_frac}.
To measure these deviations from standard physics expectations below a cutoff 
energy around 100 GeV and to verify the agreement above the cutoff 
requires precise positron spectroscopy with the dominant proton background
reduced by a factor of 10$^6$. 
Three orders of magnitude will be achieved with electromagnetic calorimeter 
shower shape measurements and a further factor of ten from matching calorimeter particle energy 
with spectrometer particle momentum.
A transition radiation detector (TRD) will provide an additional 
proton rejection factor between 100 and 1000.

\section{The AMS-02 Detector}
\label{section2}

\begin{figure}[H]
\vspace{-0.5cm}
\hspace{-0.0cm}
\includegraphics[width=7.5cm,clip]{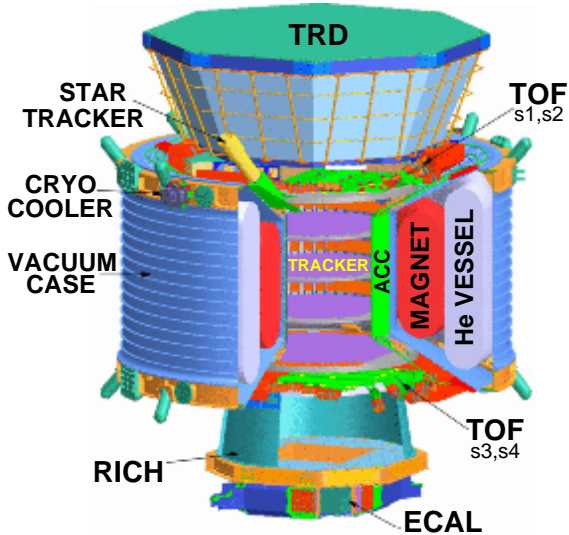}
\vspace{-1.0cm}
\caption{\it AMS-02 detector configuration}
\label{AMS02_explv}
\end{figure}

The AMS-02 experiments is 
optimised for precision particle spectroscopy in space, based on the 
experience gained from the precursor flight of AMS-01 for ten days in 1998 \cite{AMS01_Results}. 
Fig. \ref{AMS02_explv} shows the configuration of the detector.
\vspace{-0.5cm}
Built around a superconducting magnet with a bending power of 0.86 Tm$^2$ 
and 8 planes of double-sided silicon tracker with 0.5 m$^2$sr acceptance
it has two crossed scintillator planes for trigger and time of flight (TOF) measurements both above and below the tracker 
and anticoincidence scintillation counters (ACC) enclosing the side of the tracker. 
The ring imaging cherenkov counter (RICH), the electromagnetic calorimeter (ECAL) at
the bottom and the TRD on top enhance the particle identification capability.
For details refer to \cite{AMS02_Construction}.

\section{Transition Radiation Detection}

Transition Radiation (TR) photons are soft X-rays peaking around 5 keV.
With a low probability on the order of $\alpha_{em}$ TR photons are generated when a boundary 
with a change in the dielectric constant is crossed by a charged particle 
with a relativistic Lorentz factor above a threshold of around 500
and are emitted colinear with the primary particle.  
Thus primary protons up to 300 GeV can be distinguished from positrons of same momentum
in a suitable detector recording both the direct ionisation signal from the primary particle
and the TR photons after a radiator with a large number of TR boundary crossings
\cite{TR_Effect}. 

\subsection{The AMS TRD}
The AMS TRD design is based on R\&D work for ground based experiments \cite{ATLAS_TRT,HERAB_TRD}.
It uses an irregular fleece radiator and Xe/CO$_2$ filled proportional wire straw tubes.
The challenge is to build such a detector for safe and reliable operation
for three years in space.
The AMS TRD will have 20 layers, each with 20mm fleece and detector modules of 16 tubes.
The straws are made of double layer aluminised kapton foils 
and have an inner diameter of 6mm. A centered 30 $\mu$m gold plated tungsten wire
is operated at 1350V for a gas gain of 3000.
The AMS TRD covers an area of 2x2m$^2$ with 328 modules built from 16 straws each
with gastightness at the diffusion level.
The construction is described in detail in \cite{AMS02_TRD_Construction,AMS02_TRD_SpaceQuali}.

\section{TRD Prototype Beamtest} 

\begin{figure}[H]
\vspace{-0.6cm}
\includegraphics[width=7.5cm,clip]{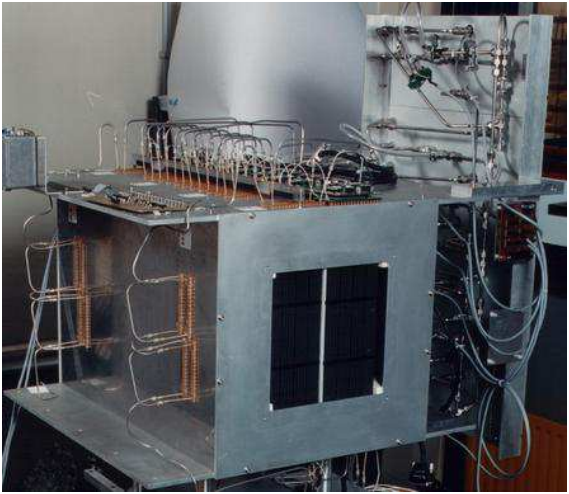}
\vspace{-1.1cm}
\caption{\it 20 layer AMS TRD prototype}
\label{trd_prototype}
\vspace{-0.6cm}
\end{figure}

To verify the proton rejection power of the AMS TRD design, 
40 modules of 40 cm length were built and arranged in two staggered towers
of 20 layers each, with layers 3, 4, 17 and 18 aligned horizontally, the others vertically
as shown in fig \ref{trd_prototype}.
The gas-panel has six separate open circuit Xe/CO$_2$ (80/20) supply lines at 1 l/h 
and atmospheric pressure for 6 or 8 modules connected in series.
The HV is set to 1480 V and is distributed from a single channel supply line.
The frontend boards are equipped with VA32\_HDR2 \cite{VA32} analog multiplexers, 
with an input range set to 1100 fC.
They are read out with an external sequencer and 12bit ADCs as were used 
for the AMS-01 tracker K-side \cite{ambrosi}. \\
Additional beamtests were performed in 2002 and 2003 with final design
frontend readout cards equipped with VA32\_HDR12 chips, onboard sequencing
and 12 bit ADCs connected to AMS TRD DAQ crate electronics
\cite{TRD_DAQ}.

\subsection{Cern Beamline and Trigger Setup}

\begin{figure}[H]
\vspace{-0.8cm}
\includegraphics[width=7.5cm,clip]{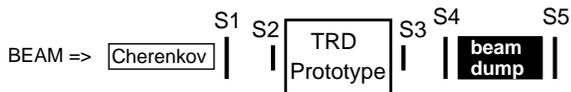}
\vspace{-1.1cm}
\caption{\it Schematic beamline trigger setup}
\vspace{-0.8cm}
\label{trigger_setup}
\end{figure}

At the CERN-SPS X7 and H6 beamlines 3~$\cdot 10^6$ events are recorded
with negative and positive charges.
Fig. \ref{trigger_setup} shows the schematic setup.
The event-trigger uses four scintillators (S1,S2,S3,S4). Two directly in front (S2) and behind (S3) the jig,
the other two at the beam entrance (S1) and in front of the beam dump (S4).
The trigger logic is set up to select a four-fold (S1\&S2\&S3\&S4) scintillator coincidence,
with a veto requiring no previous signal for at least 100 $\mu$s from
scintillators S1 or S4
to reject those particles that are closely preceeded by a previous particle crossing,
that would influence the signal height due to base-line shifts from remaining signal overshoot.\\
An additional scintillator (S5) is placed behind the beam-dump
to select or veto on muons.\\
Further particle identification is performed at the X7 beamline with cherenkov counter signals from
two threshhold-counters (He and N$_2$) each set to efficiently veto the lighter particles
when recording protons or pions.\\
At the H6 beamline a proton trigger requires the 8-fold coincidence of a CEDAR-N He differential counter 
with settings optimised according to a pressure scan for each beam energy.
At the X7 beamline electrons, muons and pions are recorded from 20 to 100 GeV, 
protons from 15 to 200 GeV (tertiary beam) and at 250 GeV (secondary beam).
At the H6 beamlines protons are recorded at 120, 160 and 200 GeV.\\
The prototype was mounted on a support allowing rotations around the vertical
axis. The data are recorded at 1.5$^\circ$ and 9$^\circ$ tilt angle between
the beam axis and the jigg.

\subsection{Data Preparation}

Prior to the physical data analysis the recorded beam data are preprocessed.
This requires calibration runs which are recorded in between the beam data runs for
pedestal subtraction and conversion from ADC bin to deposited energy.
In addition the gas pressure and temperature are continuously monitored
for gas density dependent gas gain corrections.
Finally clean single track events are preselected.

\subsubsection{Raw data preparations}

\begin{figure}[H]
\vspace{-0.8cm}
\includegraphics[width=7.5cm,clip]{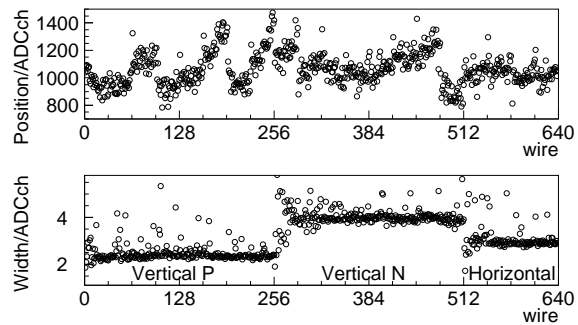}
\vspace{-1.1cm}
\caption{\it Pedestal positions and widths}
\vspace{-0.8cm}
\label{pedestal_pos_wid}
\end{figure}

\begin{figure}[H]
\vspace{-0.8cm}
\includegraphics[width=7.5cm,clip]{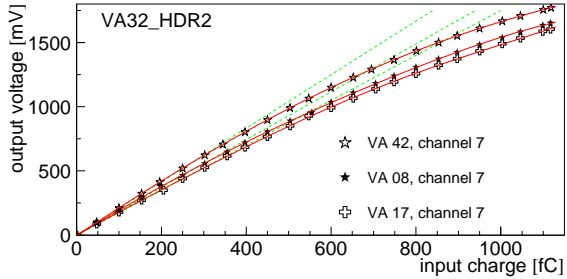}
\vspace{-1.1cm}
\caption{\it VA32\_HDR2 input response}
\vspace{-0.8cm}
\label{VA32_nonlin}
\end{figure}

Pedestal positions and widths are monitored for the complete data taking period 
and found to be stable.
They are shown in fig. \ref{pedestal_pos_wid} for all 640 channels.
With 4096 bins from a 12bit ADC the dynamic range is around 3000 bins above pedestal.
The noise between 2 and 4 ADC channels is estimated from the pedestal widths $\sigma$
after individual common mode correction for each group of 16 channels from one module. 
It varies for the three frontend boards (vertical-N, vertical-P, horizontal) 
due to the different noise conditions on different sides on the jig.
The individual pedestal positions and widths are used to define a {\it hit} as a
wire signal larger than 5$\sigma$ above pedestal.
Also the deviations from a linear input response of the VA Chips are corrected.
Fig. \ref{VA32_nonlin} shows the typical VA input response.
The improved flight module frontend boards have pedestal positions 
between 300 and 600 ADC channels, 
a linear response up to 1800 fC and the electronic noise is below 2 ADC channels.

\subsubsection{Intercalibration}

\begin{figure}[H]
\vspace{-0.8cm}
\includegraphics[width=7.5cm,clip]{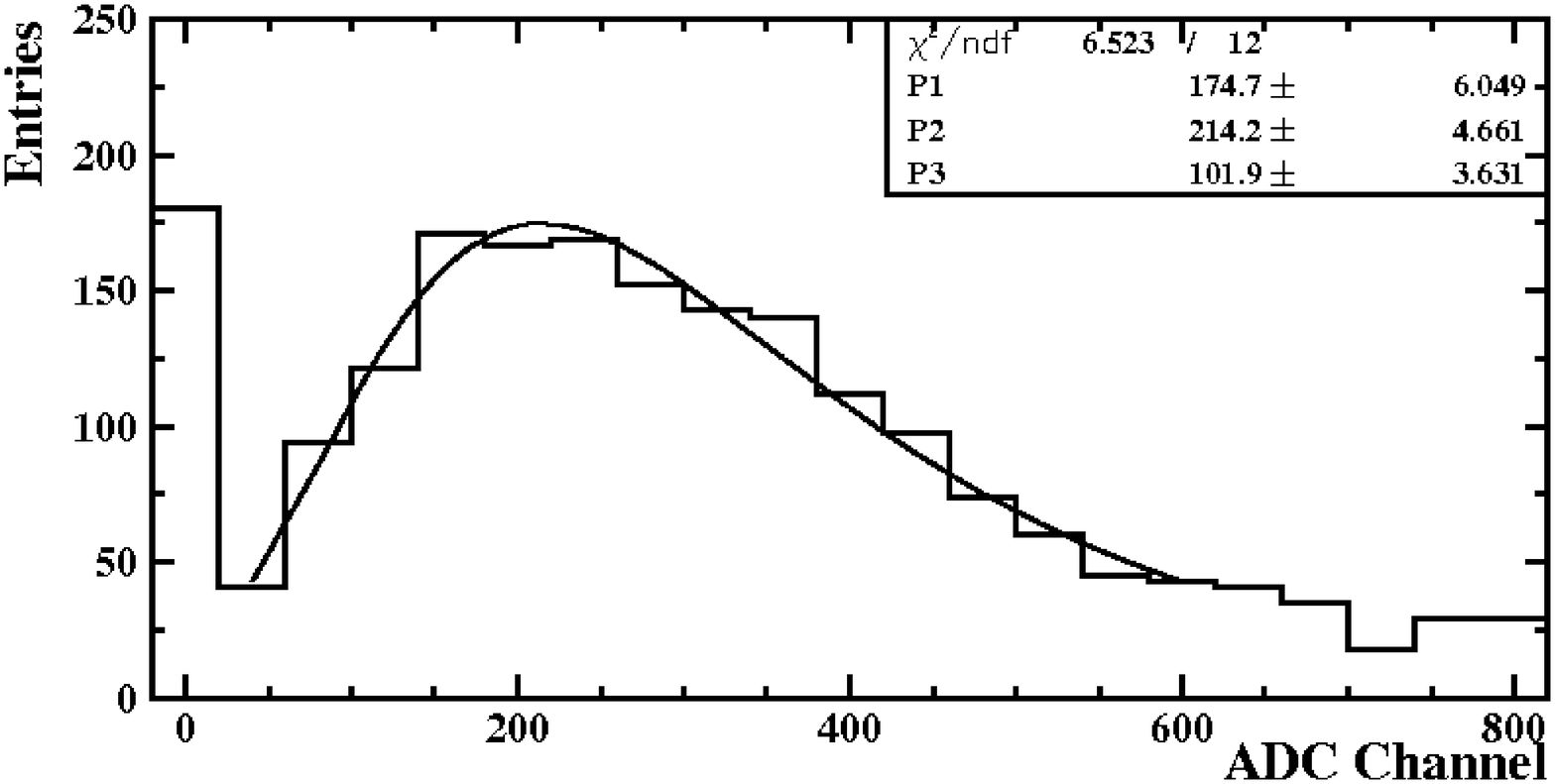}
\vspace{-1.1cm}
\caption{\it Single wire muon pulseheight distribution with pseudo-landau parameterization}
\vspace{-0.8cm}
\label{muon_etube}
\end{figure}

With 5000 muon hits per tube a relative pulse height intercalibration with errors
below 2\% is achieved. 
Fig. \ref{muon_etube} shows a typical muon pulseheight distribution used for these fits.

The intercalibration factors vary between 0.9 and 1.1, mainly due to 
VA preamplifier gain variations. The contribution from straw tube gas gain variation is below 2\%.

\subsubsection{Gas density correction}

\begin{figure}[H]
\vspace{-0.8cm}
\includegraphics[width=7.5cm,clip]{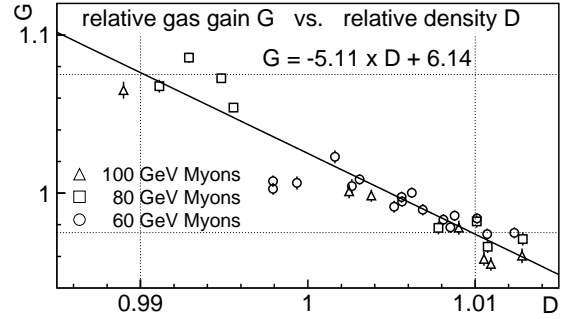}
\vspace{-1.1cm}
\caption{\it Gas gain to gas density correlation}
\vspace{-0.8cm}
\label{corr_gain_density}
\end{figure}

The muon events are also used to determine the gas gain dependence on the gas density.
Fig. \ref{corr_gain_density} shows the correlation. 
It is used to normalise the
recorded data runs to standard temperature and pressure conditions. 
At the beamtest operating parameters a 1\% increase in gas density 
results in a 5\% decrease in gas gain.

\subsubsection{Energy Calibration}

\begin{figure}[H]
\vspace{-0.8cm}
\includegraphics[width=7.5cm,clip]{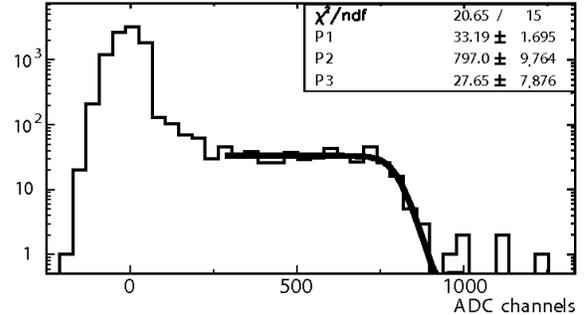}
\vspace{-1.1cm}
\caption{\it Random triggered $^{55}$Fe signal with Fermi-Function parameterization}
\vspace{-0.8cm}
\label{fe55_fit}
\end{figure}

\begin{figure}[H]
\vspace{-0.8cm}
\includegraphics[width=7.5cm,clip]{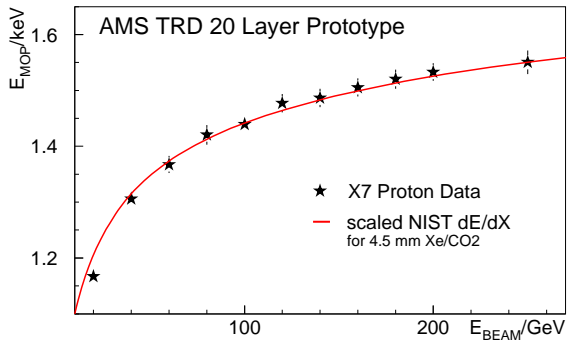}
\vspace{-1.1cm}
\caption{\it Most probable proton ionisation energy loss in 4.5mm of Xe/CO$_2$}
\vspace{-0.8cm}
\label{proton_dedx}
\end{figure}

In between runs an $^{55}$Fe source was placed in front of layers 1 and 20. 
To calibrate the ADC energy scale the upper edge of a periodically triggered spectrum 
is fitted as shown in fig. \ref{fe55_fit} and identified as the 
upper edge of the 5.9 keV photopeak.
All  $^{55}$Fe calibration measurements are averaged after tube intercalibration
and gas density corrections to give an overall energy calibration factor of
9.14 eV per ADC channel. 
For the recorded proton events fig. \ref{proton_dedx} shows the most 
probable ionisation energy loss. The beam energy dependence is in good agreement
with a globally scaled NIST prediction\cite{NIST_estar} for the relativistic rise. 

\subsubsection{Single track event selection}

\begin{figure}[H]
\vspace{-0.8cm}
\includegraphics[clip,width=7.5cm]{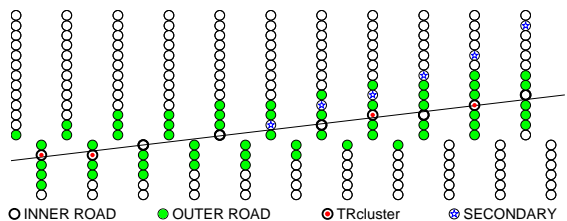}
\vspace{-1.1cm}
\caption{\it Inner and outer road around the primary track-fit}
\vspace{-0.8cm}
\label{track_inner_outer_road}
\end{figure}

With AMS good proton positron separation will be required primarily for clean unambiguous single track events.
To select these from the recorded beamtest data a track finding algorithm is used which assumes
only one track in the detector: Separately for the 16 vertical (v) and 4 horizontal (h) layers 
a first linear fit is performed with mean hit positions from each layer of
the detector using their RMS as errors to give layers with a single hit a higher weight.
A second linear fit uses the 10(v) and 4(h) hits with the smallest residuum
with respect to the first fit. 
Around this path an inner and outer road are defined containing the wires within 1.5 and 5 tube diameters
as illustrated in fig \ref{track_inner_outer_road}.
A clean single track event requires 10(v) and 3(h) hits on the inner road
with at least one hit in the first and last pair of layers, 
less than 4(v) and 2(h) hits in the outer road and
less than 6(v) and 3(h) hits in the rest of the detector.
For beamtest data the single-track efficiency varies between 50\% and 70\%
depending on the beamline settings. 
The expected single track efficiency for space station data is estimated 
from MC simulations of single particle passages. 
It is around 90\% with 5\% loss from interactions inside the TRD 
and 5\% from interactions in the material in front of the TRD.

\subsection{Selected events}

These clean single track events are selected to study particle separation and the detector simulation.

Electron data from runs above 20 GeV beam-energy were found to have
a lower clean single track event fraction.
Since the remaining event statistics do not contribute significantly,
but apart from that are in agreement with the 20 GeV electron data, 
all electron data presented here are those recorded at 20 GeV.   

The TRD Monte Carlo simulation (MC) is based on Geant3 \cite{Geant3} 
with modifications to generate and absorb TR photons \cite{TRsim}  
and to improve ionisation fluctuations in thin gas layers \cite{DEDXsim}, 
as implemented by the HERA-B collaboration \cite{Geant3TRD}.
Precomputed TR- and ionisation-dN/dX tables for primary particle 
Lorentz-factors from 1.5 to 200000 are used to randomly generate individual 
TR-photons and ionisation energy losses for each Geant tracking step. 

The tube energy deposition distributions for recorded electrons at 20 GeV and protons from 20 to 160 GeV
are used to tune the TR generation parameters to the AMS TRD radiator specifications
and the TR absorption and ionisation fluctuation parameters to the AMS TRD straw wall and gas specifications
\cite{Orboeck_PhD}. 

\begin{figure}[H]
\vspace{-0.8cm}
\includegraphics[width=7.5cm,clip]{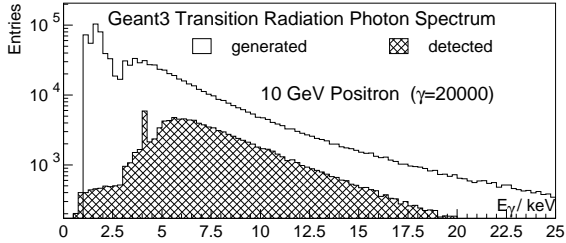}
\vspace{-1.1cm}
\caption{\it Geant3 simulation for energy spectra of generated and detected TR photons.}
\vspace{-0.8cm}
\label{trphoton_sim}
\end{figure}

With the Geant3 simulation the energy depositions from ionisation and TR photons can be studied separately.
Fig. \ref{trphoton_sim} shows the energy spectrum for all TR photons generated in the fleece radiator
and for those which are detected through absorption in the counting gas.
The detection efficiency for TR photons above 5 keV is around 30 \%.
It drops for lower energeties due to reabsorption in the radiator.
The line at 4.2 keV are Xenon L-escape photons.
  
\subsubsection{Single tube spectra}

\begin{figure}[H]
\vspace{-0.8cm}
\includegraphics[width=7.5cm,clip]{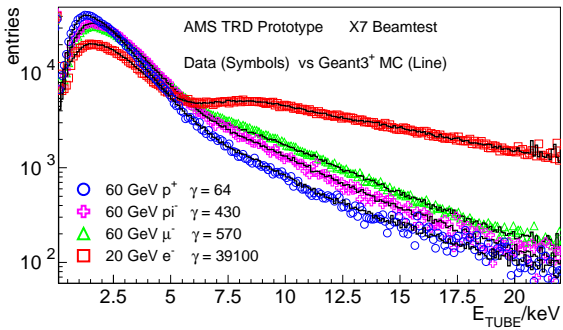}
\vspace{-1.1cm}
\caption{\it Tube energy depositions for protons, pions, muons and electrons from X7 beamtest vs. modified Geant3 simulation.}
\vspace{-0.8cm}
\label{etube_eppimu}
\end{figure}

Fig. \ref{etube_eppimu} shows the recorded energy depositions in a single tube for different
particles and different beam energies. The amount of TR contribution above 5 keV can be seen to depend on 
the Lorentz-factor $\gamma$ of the primary particle and it is well described by the modified Geant3 MC simulation. 

\section{Electron Proton Separation}

Depending on their Lorentz factor the particles will be separated based on the tube energy deposition. 
The background rejection factor is defined as the inverse of the background selection efficiency 
with cut parameters set for a signal selection efficiency of 90\%.
The recorded electron data are assumed identical to the positron distributions. 

\subsection{Cluster Counting}

\begin{figure}[H]
\vspace{-0.8cm}
\includegraphics[width=7.5cm,clip]{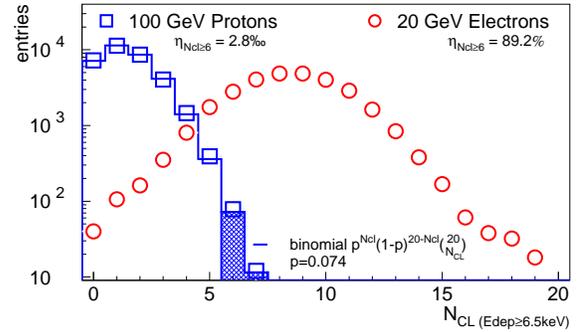}
\vspace{-1.5cm}
\caption{\it Number of clusters above 6.5 keV on a track for 20 GeV electron and 100 GeV proton X7 beamtest data} 
\vspace{-0.9cm}
\label{n_cluster}
\end{figure}

The cluster counting algorithm defines a {\it cluster} as a tube on the particle track
with a deposited energy above a fixed threshold: Edep $>$ Ethr.
Events with a cluster count Ncl $\ge$ Ncut are selected.
Fig. \ref{n_cluster} shows the number of clusters for 20 GeV electrons and 100 GeV protons
with a threshold cut of Ethr = 6.5 keV. 
The proton cluster count is well described by a binomial distribution,
reflecting the statistical nature of the ionisation fluctuations.
For Ncut = 6 the shaded area indicates a proton rejection factor of $360\pm60$.

\subsection{Likelihood\label{likeli}}

\begin{figure}[H]
\hspace{-0.5cm}
\includegraphics[width=7.5cm,clip]{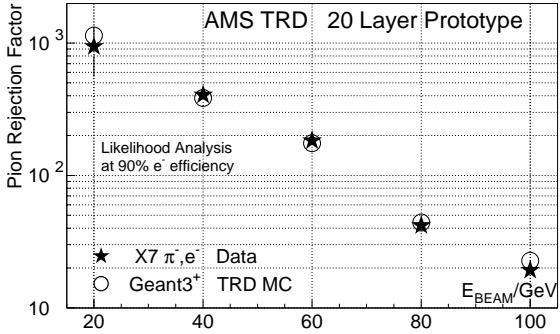}
\vspace{-1.0cm}
\caption{\it TRD beamtest pion rejections for 90\% electron efficiency}
\label{pion_rejection}
\vspace{-0.8cm}
\end{figure}

According to the Neyman-Pearson-Lemma \cite{ney_pea} the best test to decide
between two hypotheses is the liklihood ratio which can be defined as:\\
\begin{math}
L = P_e(\vec E)/(P_e(\vec E) + P_p(\vec E))
\end{math} 
where $\vec E$ is the vector with the energy deposition of one event.\\
$P_{\mbox e}(\vec E)$, $P_{\mbox p}(\vec E)$ are the multidimensional
probabilities for this event to be electron- or protonlike. 
To reduce the dimensions of this problem mean probabilities 
$\bar P_{\mbox e}$, $\bar P_{\mbox p}$ are used instead of 
$P_{\mbox e}(\vec E)$, $P_{\mbox p}(\vec E)$ to classify the events:
\begin{math}
\bar P_{\mbox{e/p}}=\sqrt[\displaystyle n]{\prod_i^n P_{\mbox{e/p}}(E^{(i)})}
\end{math}\\
where $n$ is the number of hits on the track.\\ 
$P_{\mbox{e/p}}(E^{(i)})$ is the probability 
for the $i$-th hit with energy $E^{(i)}$ to be electron- or protonlike. 
It is determined from a parameterisation to the histograms of the tube energy deposition, 
averaged over all layers. 
For clean single track events this likelihood analysis achieves proton rejection factors 
from 2000 to 100 for proton energies from 20 GeV to 250 GeV.
To allow a comparison with other TRD experiments, the pion-electron separation power is shown in fig.\ref{pion_rejection}.  
\vspace{-0.8cm}

\subsection{Neural Network}

\begin{figure}[H]
\hspace{-0.5cm}
\includegraphics[width=7.5cm,clip]{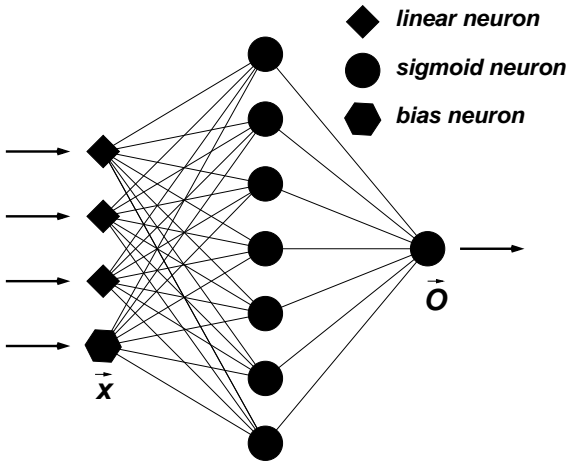}
\vspace{-0.8cm}
\caption{\it Neural network with one hidden layer}
\label{nn}
\vspace{-0.5cm}
\end{figure}

To further improve the positron proton separation power, the beamtest data are
analysed with a neural network algorithm \cite{fisher}. A neural network adapts its 
weights with the aim to map training events to expected values (learning). 
The feedforward neural net used here consists of three layers with one neuron
in the output layer. 
In the hidden and the last layer neurons with a sigmoid output function 
$o_i = \tanh A_i$ are used (fig. \ref{nn}) because a differentiable output
function is needed 
for the learning algorithm. $A_i$ is the activation of the particular neuron.
\vspace{-0.6cm}

To determine the goodness of the mapping of the input data to the expected
values, the mean square error (MSE) is defined as:
\begin{equation*}
F(\vec w) = \lim_{N\rightarrow\infty}\frac1N\sum_{k=1}^{N}\underbrace{\left(\vec  E^k(\vec x_k) 
          - \vec o^k(\vec x_k,\vec w\right)^2}_{\displaystyle f_k},
\end{equation*}
where $N$ is the number of training events, $\vec w$ is the vector with all weight values, 
$\vec E^k$ is the expected value and $\vec o^k$ the actual value for the $k$-th event.

The weight adaption is done with the $\bar\delta$-$\delta$-learning rule  
which is an extension of the backpropagation algorithm
that adapts not only the weights but also the learning rates for each weight on the basis of the MSE:
\begin{equation*}
\begin{split}\Delta w_i(t+1) &= \eta_i(t+1)\cdot\frac1N\sum_{k=1}^N\partial_{w_i}f_k^{t+1}\\
                             &+ \alpha\cdot\Delta w_i(t),\end{split}
\end{equation*}
\begin{equation*}\footnotesize
\eta_i(t+1)=\begin{cases} \eta_i(t)\cdot(1-\eta^-), & \partial_{w_i}f_k^{t+1}\cdot\partial_{w_i}f_k^{t}<0, \\ 
\eta_i(t)+\eta^+, & \partial_{w_i}f_k^{t+1}\cdot\partial_{w_i}f_k^{t}>0.\end{cases}
\end{equation*}
\normalsize $\Delta w_i(t)$ is the change of weight $i$ in step $t$ and $\eta_i(t)$ 
is the learning rate at this step. $\partial_{w_i}f_k^{t}$ is the error gradient 
of event $k$ at step $t$. $\alpha$, $\eta^+$ and $\eta^-$ are free parameters 
for the optimization of the learning process. The weight update takes place 
after each presentation of a batch with 200 training events. 
A random presentation prevents a learning of the batch order.
This algorithm achieves good results with low CPU costs.

A random initialization of net weights according 
to a normal distribution with $\sigma_{\mbox w} =1/\sqrt{n}$, 
where $n$ is the number of inputs to a layer, is a good starting point 
for the learning process \cite{effback}.  
To present all inputs to the net equally, they are normalised 
for each input neuron with $x_k^{\prime}=(x_k-\bar x)/\sigma$.
The mean value $\bar x$  and the width $\sigma$ 
are calculated from the training sample for each input neuron.

The complete event sample is randomly divided into three subsamples: 
40\,\% for training, 20\,\% for validation during training 
and the remaining 40\,\% for the "´real-world"´ test data 
to determine the proton rejection factor.
The training will be stopped when the MSE of the validation sample 
reaches its global minimum. 

\subsubsection{Tube Energy Neural Network}

\begin{figure}[H]
\hspace{-0.5cm}
\includegraphics[width=8.5cm,clip]{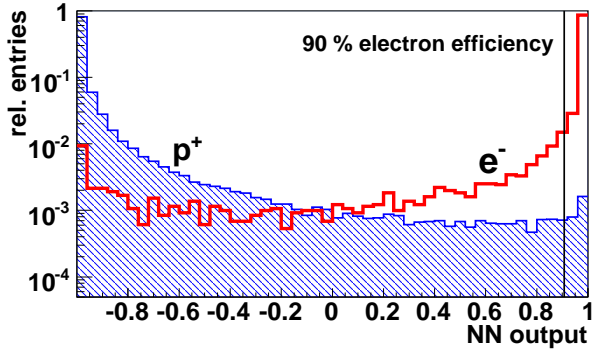}
\vspace{-1.2cm}
\caption{\it Trained network output for protons (hatched) and electrons.}
\label{output}
\vspace{-0.4cm}
\end{figure}

\begin{figure}[H]
\hspace{-0.5cm}
\begin{minipage}[b]{.4\linewidth}
\includegraphics[width=4cm,clip]{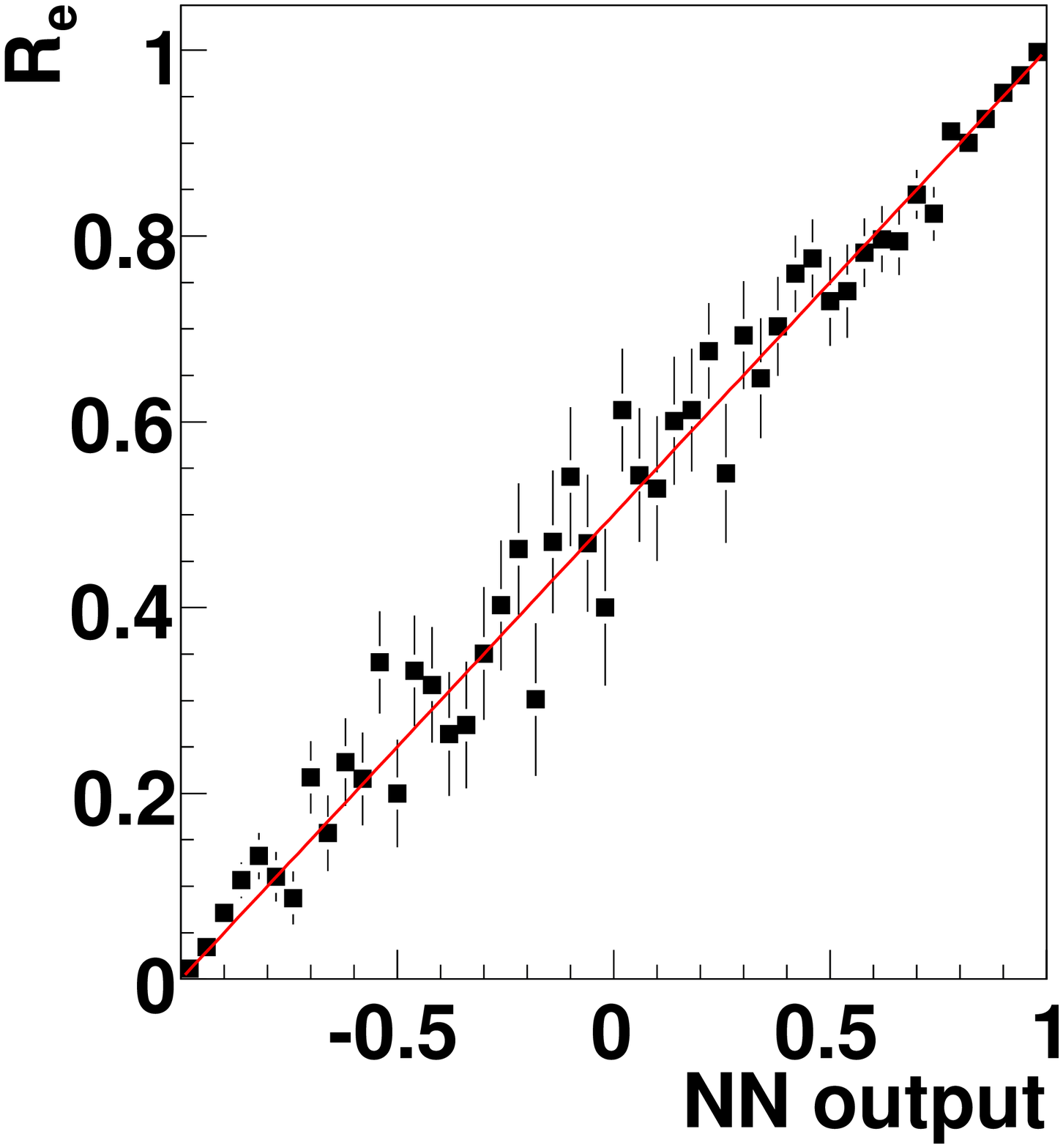}
\end{minipage}
\hspace{.1\linewidth}
\begin{minipage}[b]{.4\linewidth}
\includegraphics[width=4cm,clip]{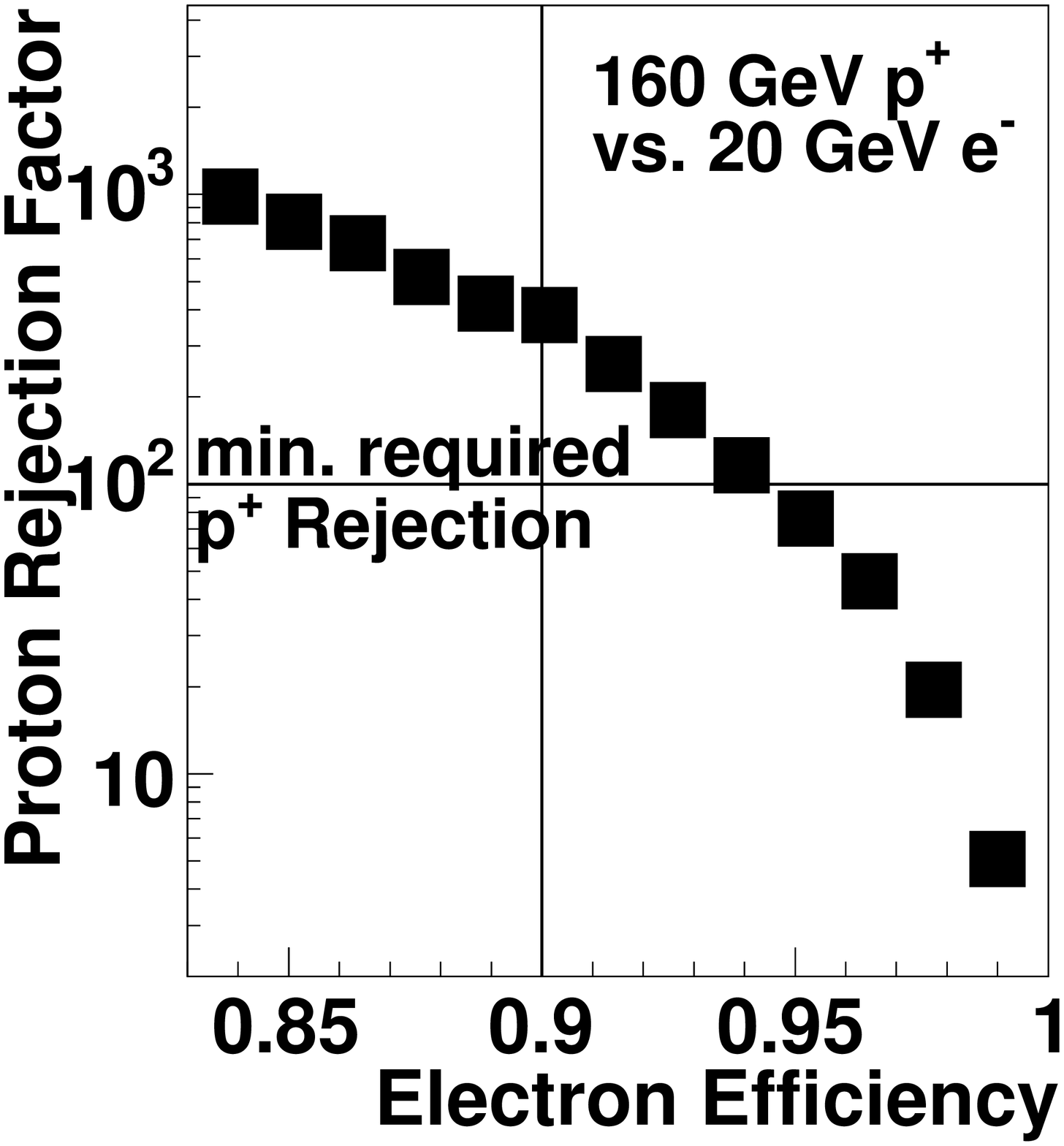}
\end{minipage}
\vspace{-1.0cm}
\caption{\it Left: Signal to background ratio $R_e$; right: Rejection to efficieny correlation.}
\label{sb_eff}
\end{figure}

In a first step the 20 energy depositions on the track are used as input data
for a neural net ($E_{\mbox{Tube}}$-NN). 
The expected value for proton events is set
to -1 and for electron events to +1. The input data consists of all single track
events at all recorded energies and tilt angles. 
This has the advantage that only one net is needed for the analysis of all
events and the statistics for the training are higher. 
After parameter variation highest rejections were achieved with
$\eta^+=5.5\cdot10^{-3}$, $\eta^-=0.6$ , $\alpha=0.6$ and 60 neurons in the
hidden layer. The output of the trained net for 60\,GeV protons and 20\,GeV
electrons is shown in fig. \ref{output}.
\vspace{-0.8cm}
For an optimally trained net the MSE 
$F_o(o) = R_{\mbox{e}}(1-o)^2+(1-R_{\mbox{e}})(-1-o)^2$
for each particular output $o$ should be minimal \cite{feindt}.
$R_{\mbox{e}} = \eta_{\mbox e}/(\eta_{\mbox e}+\eta_{\mbox p})$ 
is the signal to background ratio of the electron $\eta_{\mbox e}$ 
and proton events $\eta_{\mbox p}$ at output $o$. 
From $\partial F_o(o)/\partial o = 0$ or $R_e = (o+1)/2$
it follows that $R_{\mbox{e}}$ is a straight line for an optimally trained net. 
On the left side of fig. \ref{sb_eff} it is shown that the net is well trained.
Variations of the cut parameter show a smooth correlation beteewn rejection and
efficiency (fig. \ref{sb_eff} right).

The AMS TOF measurement will provide the particle direction. 
Apart from a slight saturation of the tube energy spectra
within the first few layers of the electron sample
the likelihood method is insensitive to the input order. 
The neural net on the other hand has an intrinsic sensitivity 
because during training every track is presented in the right order
of the particle energy depositions in the detector. 
A classification run with the energy depositions given in random or reversed order
shows a rejection factor reduction of 25 \%. 

\begin{figure}[H]
\hspace{-0.5cm}
\includegraphics[width=8.5cm,clip]{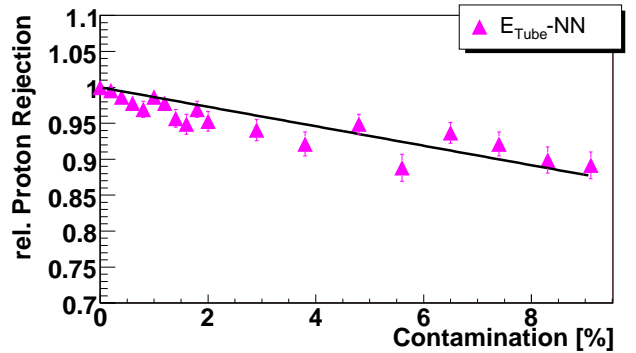}
\vspace{-1.5cm}
\caption{\it Proton rejection with contamination of training and validation
             sample for $E_{\mbox{Tube}}$-NN}
\label{cont}
\vspace{-0.4cm}
\end{figure}

The AMS TRD training sample will eventually consist of cosmic electrons and
protons preselected with the AMS ECAL.
To test the dependence on the input data sample purity, the electron
training and validation samples are contaminated with a fraction of
protons. The normalised rejections in fig. \ref{cont} show that even a
contamination of the samples with a few percent results in rejections of the
same order as without contamination. With cosmic data a contamination of the 
training and validation samples below a permil is expected.

\vspace{-0.4cm}

\subsubsection{Combined Neural Net}

\begin{figure}[H]
\hspace{-0.5cm}
\includegraphics[width=8.5cm,clip]{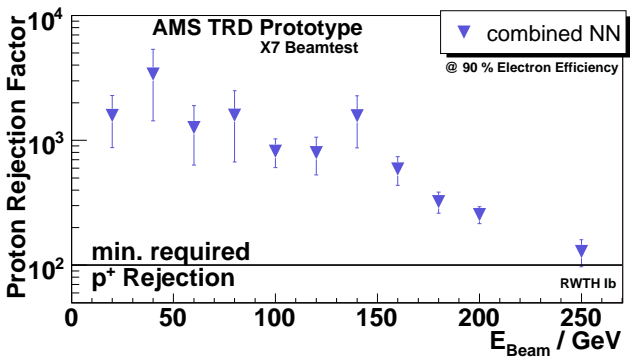}
\vspace{-1.3cm}
\caption{\it Proton rejection factors at 90\% electron efficiency for the combined neural net}
\label{all}
\vspace{-0.8cm}
\end{figure}

In a second step an additional neural network with different input data is
used. The singular analysis techniques are not the best hypotheses tests
according to the Neyman-Pearson-Lemma as pointed out in section \ref{likeli}, so
it is possible to get better results with a combination of the different methods
in an additional neural network (combined NN). A correlation analysis was
performed to find the inputs with the lowest correlation, so the combined NN can
benefit from non redundant information. The following four inputs have a mean
correlation coefficient of $|\bar r|\approx0.7$ among each other:
\vspace{-0.2cm}
\begin{itemize}\setlength{\itemsep}{-0.1cm}
\item $E_{\mbox{Tube}}$-NN,
\item Likelihood with probability distributions averaged over all beam energies.
\item TR cluster count with E$_{THR}$ = 6.5\,keV,
\item Fisher's discriminant with mean covariance matrix 
      calculated from all events.
\end{itemize}
\vspace{-0.2cm}
Every event is characterised by these four values. A neural net with 8 neurons in the
hidden layer and $\eta^+=5.5\cdot10^{-3}$, $\eta^-=0.6$, $\alpha=0.9$ gives
the best rejections. 
The rejection power of the combined net 
is shown in fig. \ref{all}. 
It achieves rejection factors above 100 over all beam energies. 
\vspace{-1.0cm}

Table \ref{compare} shows a comparison of the rejection factors for the test beam data
averaged over all recorded energies for the different classification techniques
normalised to the cluster counting method. 
The rejection of the likelihood method with mean probabilities is lower than with
individual probabilities, but the rejections of the combined net are an
improvement. The neural net with energy inputs has somewhat lower rejections than
the likelihood but a much higher rejection than the cluster counting and Fisher's
discriminant. A large fraction of the gain in separation power of the combined net over
the likelihood method is attributed to the directional information used in the
tube energy neural net.
\vspace{-0.6cm}
\begin{table}[H]
\caption{\it Comparison of beamtest proton rejections at 90\% electron efficiency for the different classification techniques}
\begin{center}
\begin{tabular}{l|c}
Method			& Proton Rejection\\
\hline
combined NN		& 2.51 $\pm$ 0.23\\
Likelihood$^1$		& 1.97 $\pm$ 0.11\\
Likelihood$^2$		& 1.87 $\pm$ 0.11\\
$E_{\mbox{Tube}}$-NN	& 1.64 $\pm$ 0.12\\
Fisher's discriminant	& 1.01 $\pm$ 0.05\\
Cluster counting	& 1.00 $\pm$ 0.05
\end{tabular}
\end{center}
$^1$ individual probabilties for each beam energy; probabilities averaged over all layers

$^2$ mean probabilties for all beam energies; individual probabilties for each layer
\label{compare}
\vspace{-1.0cm}
\end{table}

The experimentally determined rejection factors should be understood 
as a lower bound for the TRD when operated on the ISS for the following reason:
For beam energies above 200 GeV 
the Geant3 simulation predicted higher proton rejection factors than observed in the data. 
Such a discrepancy was not observed for pions and muons with Lorentz factors above 200.
As a conservative approach this was attributed to quasi elastic low energy pion pair production 
by high energy protons in the detector -- resembling a TR signature -- 
and was added to the simulation code. 
If on the other hand this effect is caused by a pion contamination present in the high energy proton beam with inefficient veto
from the cherenkov counters this  would not affect measurements on the ISS where no primary pion component is present.
Also a pion or muon contamination in the negative electron beam would reduce the measured electron efficiency
thus also underestimate the TRD performance on the ISS.

\section{Sensitivity to TRD Operating Parameters}

Uncertainties in the positron efficiency and proton suppression will 
enter as systematical errors into any positron or anti proton analysis. 
In the following analysis the tube energies are deliberately smeared 
to estimate the effects on proton rejection factors and electron efficiencies,
keeping the selection procedures and cuts fixed.
 
Electronic noise is simulated by adding a normal distributed random value 
with a relative width (rel.Noise)  and an absolute width (abs.Noise) 
to the original energy depositions.
To estimate the effect of calibration precision, 
a normal distributed set of intercalibration factors for each individual tube
is generated with a fixed width around a mean of 1.
The effect of a global change in gas gain is estimated
with a correlated variation of the energy depositions using a fixed scale factor.

\begin{figure}[H]
\hspace{-0.5cm}
\begin{minipage}[b]{.4\linewidth}
\includegraphics[width=4cm,clip]{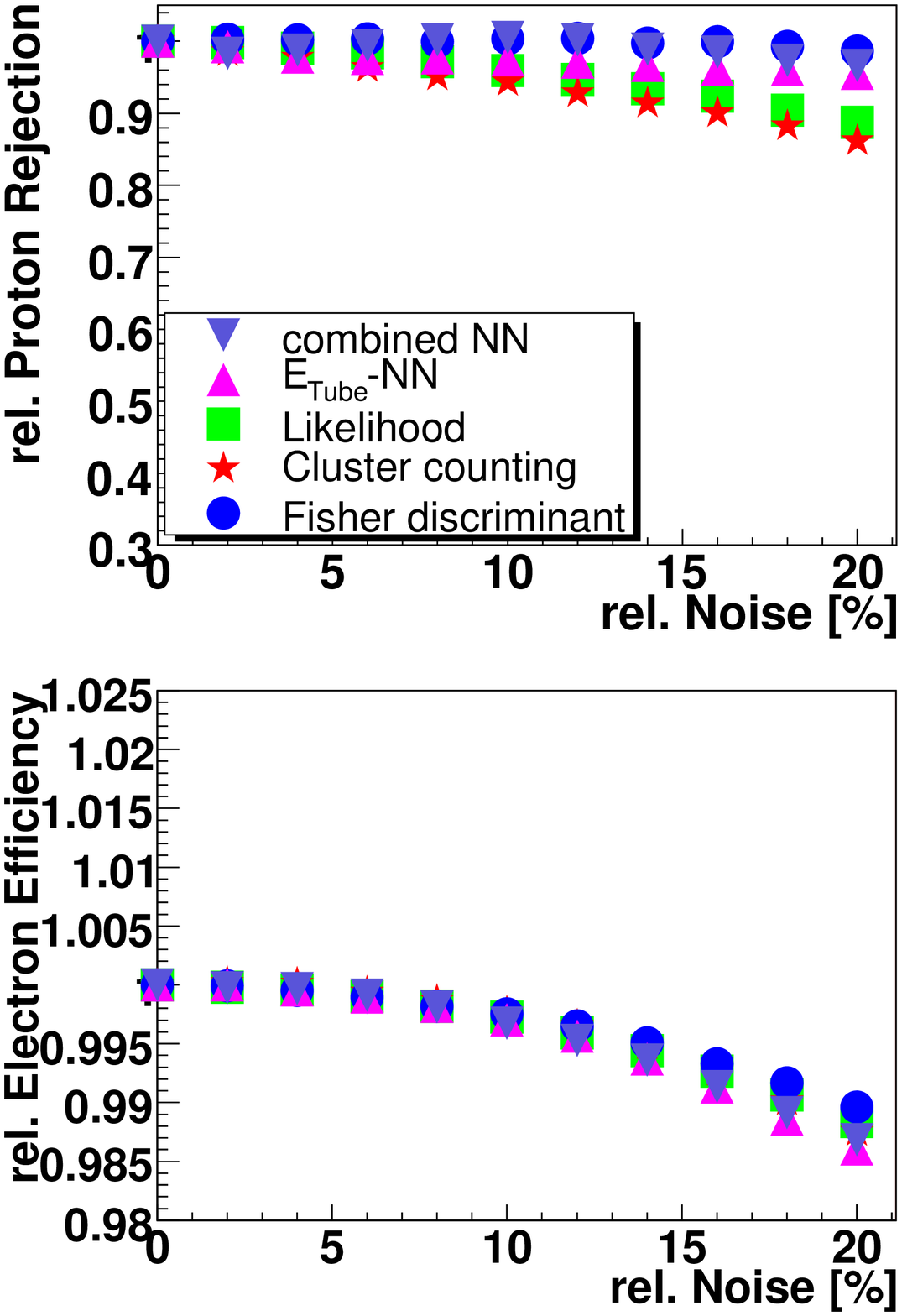}

\bigskip

\includegraphics[width=4cm,clip]{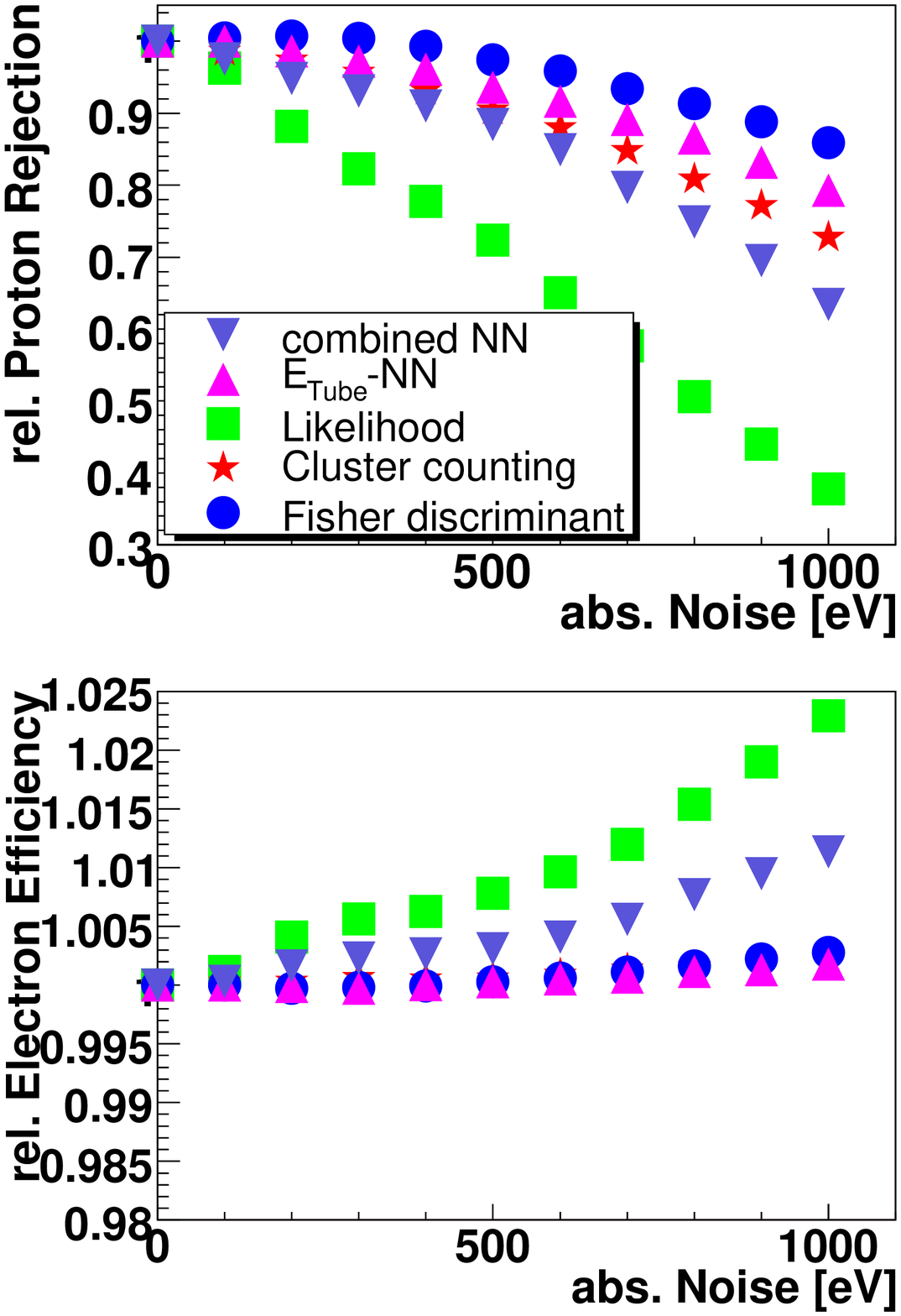}
\end{minipage}
\hspace{.1\linewidth}
\begin{minipage}[b]{.4\linewidth}
\includegraphics[width=4cm,clip]{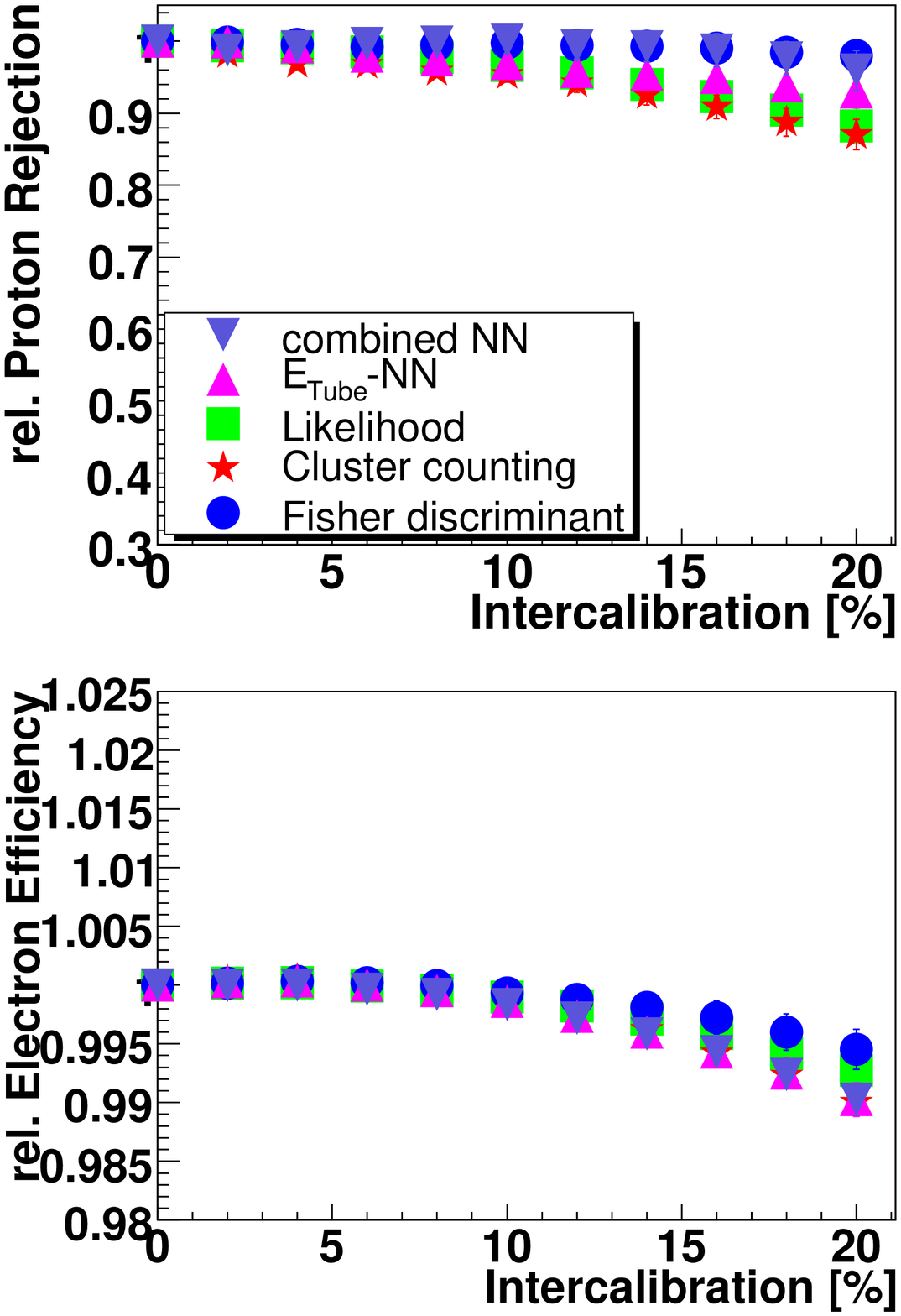}

\bigskip

\includegraphics[width=4cm,clip]{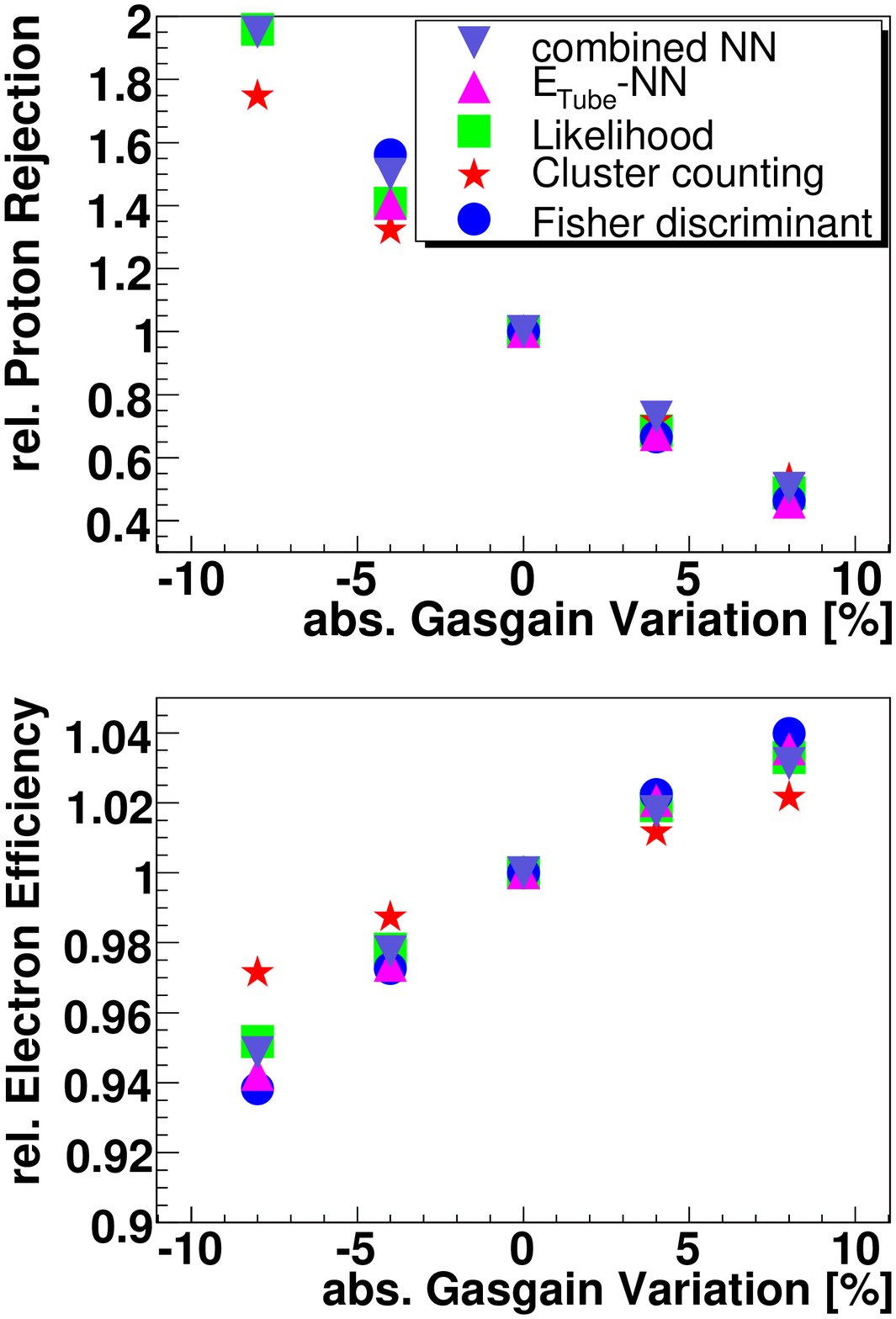}
\end{minipage}
\vspace{-0.4cm}
\caption{\it Simulation of relative variations of proton rejection factors and electron
             selection efficiencies. 
             Relative electronic noise (top left); 
             Absolute electronic noise (bottom left); 
             Intercalibration precision (top right); 
             Gas gain variation (bottom right).}
\label{noise_gg}
\vspace{-0.4cm}
\end{figure}

Fig. \ref{noise_gg} shows that the analysis techniques react quite similar to
the different kinds of smearing. The simulations show only a small dependance on
relative noise and intercalibration errors.
The likelihood method shows a larger dependance on absolute noise contamination
because the single tube probability functions are more sensitive to energy
shifts than the other methods which rely directly on the tube energies. 
With an expected electronic noise of about 2 ADC bins or 50\,eV all
methods are quite stable though.\\
The strongest effect is observed for a correlated variation of the energy depositions:
A 5\% change in gas gain -- resulting from a 1\% change in gas density --
leads to a 50\% change of the proton rejection factor, emphasizing the need
for precise gas quality control and monitoring.

\section{Conclusions}

The AMS-01 precursor flight demonstrated the feasibility  to operate a
modern particle physics detector in space, 
collecting 10$^8$ events up to rigidities of 140 GV.
To exploit AMS-02 data with statistics a 100 times higher, 
improved particle identification is required.
Beam test results have shown that a 20 layer TRD with fleece radiator
and Xe/CO$_2$ straw tube detection will deliver proton rejection factors
between 1000 and 100 for particle momenta from 20 GeV to 250 GeV.
The TRD flight version is under construction
and so far within specifications and schedule.
This project is funded by the German Space Agency DLR under contract No. 50OO0501, 
the US Department of Energy DOE and NASA.

\section*{Acknowledgments}
The authors wish to thank the follwing people 
for their support at Cern during data taking:
A.~Arefiev, V.~Pojidaev (ITEP), 
P.~Berges, J.~Tsai, M.~Vergain (MIT), 
L.~Gatignon (Cern, X7 Beamline), 
K.~Elsener (Cern, H6 Beamline)

\end{document}